\documentstyle[preprint,aps,prb,floats]{revtex}

\newcommand{\ee}{\mbox{{\boldmath $\varepsilon$}}}

\newcommand{\eaa}{\mbox{{\boldmath $\varepsilon$}$\parallel \!\! {a}$}}
\newcommand{\ebb}{\mbox{{\boldmath $\varepsilon$}$\parallel \!\! {b}$}}
\newcommand{\ecc}{\mbox{{\boldmath $\varepsilon$}$\parallel \!\! {c}$}}

\newcommand{\kedge}{{\it K} edge}

\newcommand{\cugeo}{\mbox{CuGeO$_{3}$}}
\newcommand{\lacuo}{\mbox{La$_{2}$CuO$_{4}$}}

\newcommand{\prevb}[1]{Phys. Rev. B {\bf #1}}
\newcommand{\prevl}[1]{Phys. Rev. Lett. {\bf #1}}

\newcommand{\ssc}[1]{Solid State Commun. {\bf #1}}
\newcommand{\jpcm}[1]{J. Phys.: Condens. Matter {\bf #1}}

\newcommand{\ea}{{\it et al}}
\newcommand{\mm}[1]{\mbox{$#1$}}
\newcommand{\bbt}[1]{\bibitem{#1}}

\tighten

\RequirePackage[dvips]{graphics}

\begin{document}

\draft

\title{Polarized x-ray absorption spectra of \cugeo\ \\ 
at the Cu and Ge {\it K}\ edges}

\author{ O.\ \v{S}ipr and A.\  \v{S}im{\accent23 u}nek}
\address{Institute of Physics, Academy of Sciences of the Czech
Republic, \\
Cukrovarnick\'{a}~10, 162~53~Praha~6, Czech~Republic}

\author{S.\ Bocharov}
\address{ Hasylab at DESY, Notkestr.\ 85, D-22603 Hamburg, Germany }

\author{G.\ Dr\"{a}ger}
\address{Fachbereich Physik der Martin-Luther-Universit\"{a}t
Halle-Wittenberg, \\
Friedemann-Bach-Platz~6, D-06108 Halle, Germany, \\ 
\mbox{ } }

\date{August 14, 2002}

\maketitle

\begin{abstract}
Polarized x-ray absorption near edge structure (XANES) spectra at both
the Cu and the Ge $K$-edges of \cugeo\ are measured and calculated
relying on the real-space multiple-scattering formalism within a
one-electron approach.  The polarization components are resolved not
only in the unit cell coordinate system (\eaa, \ebb, \ecc) but also in
a local frame attached to the nearest neighborhood of the
photoabsorbing Cu atom.  In that way, features which resist a
particular theoretical description can be identified.  We have found
that it is the out-of-CuO$_{4}$-plane $p_{z'}$\ component which defies
the one-electron calculation based on the muffin-tin potential.  For
the Ge \kedge\ XANES, the agreement between the theory and the
experiment appears to be better for those polarization components
which probe more compact local surroundings than for those which probe
regions with lower atomic density.\\
{\em Paper published in Phys. Rev. B {\bf 66}, 155119 (2002) and available
on-line at http://link.aps.org/abstract/PRB/v66/e155119.}
\end{abstract} 



\narrowtext

\section{Introduction}   \label{intro}
 
The main incentive for studying \cugeo\ comes from the fact that it
exhibits the spin-Peierls transition.\cite{hase} Apart from that, it
is a member of the very interesting family of copper oxides, which
displays a manifold of outstanding physical properties.  X-ray
absorption spectroscopy (XAS) is a convenient tool for investigating
low-lying unoccupied electron states, due to its chemical and angular
selectivity. This is especially important for compounds, where by
exploring different absorption edges, XAS is able to provide a
comprehensive picture of the local electronic structure of the
material in question. Further details can be obtained by analyzing
polarized (i.e., angular-dependent) spectra, as in that way features
which would be obscured in unpolarized spectra can be observed.  Also,
comparison with polarized experimental x-ray absorption spectra poses
a much more stringent test to the theory.\cite{v2o5}

\cugeo\ has an orthorhombic crystal structure\cite{brauninger} with
$a$=4.80~\AA, $b$=8.48~\AA, and $c$=2.94~\AA\ (space group no.\ 51,
labeled {\it Pbmm} or D$^{5}_{2h}$), therefore, its polarized spectra
generated by dipole transitions can be decomposed into three partial
spectral components.\cite{brouder} While all Cu and all Ge atoms are
equivalent within the symmetry properties of the lattice, there are
two distinct O sites in the \cugeo\ crystal.  Instructive pictures of
the \cugeo\ structure can be found in Fig.\ 1 of Ref.\
\onlinecite{parmigiani}, Fig.\ 1 of Ref.\onlinecite{atzkern} or, from
a more local point of view, in Fig.\ 1 of Ref.\
\onlinecite{zagoulaev}.

In the past, x-ray absorption near-edge structure (XANES) of the \ebb\
and \ecc\ polarized components at the Cu \kedge\ was measured by Cruz
\ea\cite{cruz} and polarized O \kedge\ spectra were measured by
Corradini \ea.\cite{corradini} No measurement of the Ge \kedge\ XANES
has been reported so far.  Cruz \ea\cite{cruz} perform a theoretical
analysis of the pre-peak relying on their many-body cluster
calculation, Corradini \ea\cite{corradini} compare the lowest in
energy part of their spectra with densities of states around oxygen
atoms as provided by many-body calculations of Villafiorita
\ea.\cite{villafiorita} No attempts to calculate the spectra in the
whole XANES range were made in either Ref.\ \onlinecite{cruz} or Ref.\
\onlinecite{corradini}.  Relying on quantitative arguments and on
analogies with spectra of other compounds, Cruz \ea\cite{cruz} suggest
a many-body interpretation of the main Cu \kedge\ XANES peak
splitting.

Due to the presence of strong electron correlations in \cugeo, a
one-particle approach apparently cannot be fully trusted for
describing the states at the bottom of the conduction band (pre-edge
region, photoelectron energy \mm{E \lesssim 5}~eV above the
threshold).  It may be worthwhile, however, to apply the one-particle
formalism based on the local density approximation (LDA) to the rest
of the XANES region.  One would be then able to complement qualitative
and empirical arguments, which have been applied for interpreting
\cugeo\ spectra in this energy domain so far, with a material-specific
{\em ab-initio} approach.

In our study, we investigate polarized XANES spectra at the Cu \kedge\
and the Ge \kedge, both experimentally and theoretically.  By
recording the spectra for several orientations of the crystal with
respect to the x-ray polarization vector \ee, we are able to explore
not only the \ebb\ and \ecc\ polarizations analyzed by Cruz
\ea\cite{cruz} but also the remaining \eaa\ component.  Our
theoretical treatment is based on the multiple-scattering formalism in
the real space.  For a more comprehensive insight into the physical
nature of the electron states probed by the Cu \kedge\ XANES, we
resolve the polarization components not only in the unit cell
coordinate system but also in a local reference frame attached to the
coordination polyhedron of the photoabsorbing Cu atom.  In that way,
features which defy a particular theoretical description can be
specifically identified.  As we include also the Ge edge into our
study, a complete view of the structure of unoccupied electron states
in \cugeo\ as seen from the copper, germanium, and
oxygen\cite{corradini} sites emerges finally.

\section{Results}

\subsection{Experiment}

\begin{figure}
\includegraphics[10mm,2mm][115mm,95mm]{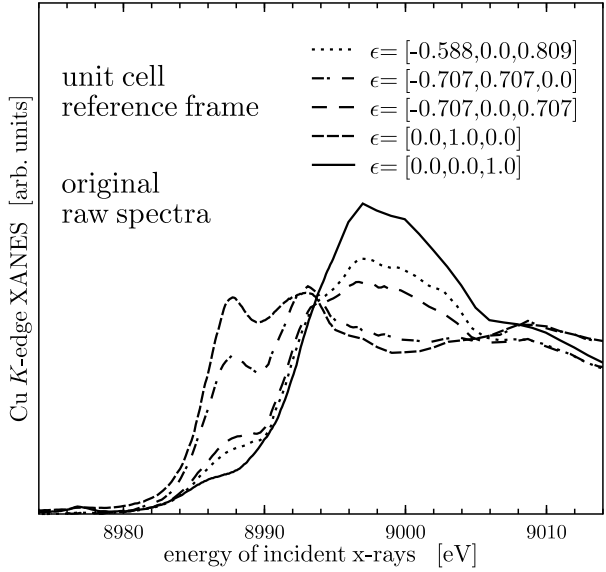}
\caption{ Measured XANES curves at the Cu edge of \cugeo\ for several
representative orientations of the polarization vector \ee\ with
respect to the \cugeo\ monocrystal.  The projected components of the
\ee\ vector on the crystal unit cell axes $a$, $b$\ and $c$ are
written in square parentheses in the legend.
\label{raw_cu}
}
\end{figure}

\begin{figure}
\includegraphics[10mm,2mm][115mm,95mm]{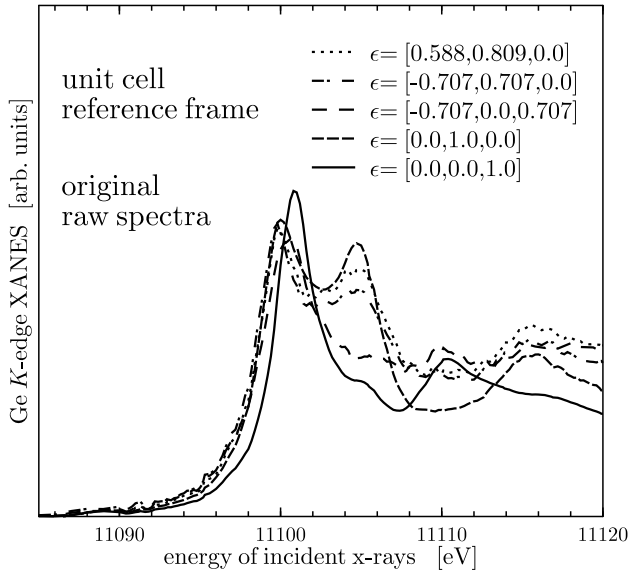}
\caption{ Measured XANES curves at the Ge edge of \cugeo\ for several
representative polarization vectors.  This drawing is analogous to
Fig.\ \protect\ref{raw_cu}.
\label{raw_ge}
}
\end{figure}

We used an experimental setup very similar to that described in a
greater detail in our previous publications.\cite{v2o5,nasecuo} The
layered structure of \cugeo\ allows to prepare a sample for x-ray
transmission experiments through splitting a single-crystal slab into
thin plates perpendicular to the unit cell direction $a$.  A plate of
about 7~mm$\times$4~mm$\times$0.03~mm was separated from the slab by
means of usual scalpel and immediately used as the sample.  The
experiments were carried out at the beam lines A1 and E4 (HASYLAB,
DESY) equipped with an Si (111) two crystal monochromator.  For
detection of the x-ray intensities in front of and behind the sample
plate, usual ionization chambers were used.  The sample plate was
positioned in a PC-controlled goniometer, allowing three perpendicular
rotations.  In that way, several polarized spectra for various
orientations of the polarization vector \ee\ with respect to the
sample were recorded.  In order to separate the absorption resulting
from the $K$-transitions exclusively, we used a subtraction of the
background function \mm{f(E) = a/E^{4} + b/E^{3} +c} after
Victoreen. All the spectra were corrected for an equivalent effective
sample thickness, with an additional matching normalization of {\it
max}$\pm 5$\% in the remote extended x-ray absorption fine structure
(EXAFS) region.  Some of the measured curves are displayed in Fig.\
\ref{raw_cu} (Cu edge) and Fig.\ \ref{raw_ge} (Ge edge) for several
representative orientations of the polarization vector \ee\ with
respect to the \cugeo\ monocrystal.

\subsection{Symmetry decomposition}

Theoretical foundations needed for extracting symmetry-resolved
partial spectral components of x-ray absorption spectra were outlined
by Brouder\cite{brouder} in an extensive way.  For an orthorhombic
crystal, in particular, any polarized x-ray absorption spectrum can be
decomposed into a weighted sum of three independent partial spectral
components (in the dipole approximation).  These partial spectral
components can be chosen so that they correspond to the polarized
spectra recorded with the \ee\ vectors parallel to each of the crystal
axes.  Hence, the dipole part of any linearly polarized spectrum can
be written as\cite{v2o5,nasecuo}
\begin{equation}
\mu_{D} \: = \: p_{x} \varepsilon_{x}^{2} \, + \, p_{y}
\varepsilon_{y}^{2}  \, + \, p_{z} \varepsilon_{z}^{2} \; \; \; ,
\label{polar}
\end{equation}
where $\varepsilon_{x}$, $\varepsilon_{y}$, and $\varepsilon_{z}$\ are
cartesian components of the polarization vector \ee\ and $p_{x}$,
$p_{y}$, and $p_{z}$\ denote the partial spectral components
(their designation reflects the corresponding projected density of
states).  Note that our Eq.\ (\ref{polar}) can be transformed into the
tensor form presented by Brouder\cite{brouder} by simple algebraic
manipulations. 

\begin{figure}
\includegraphics[50mm,110mm][145mm,230mm]{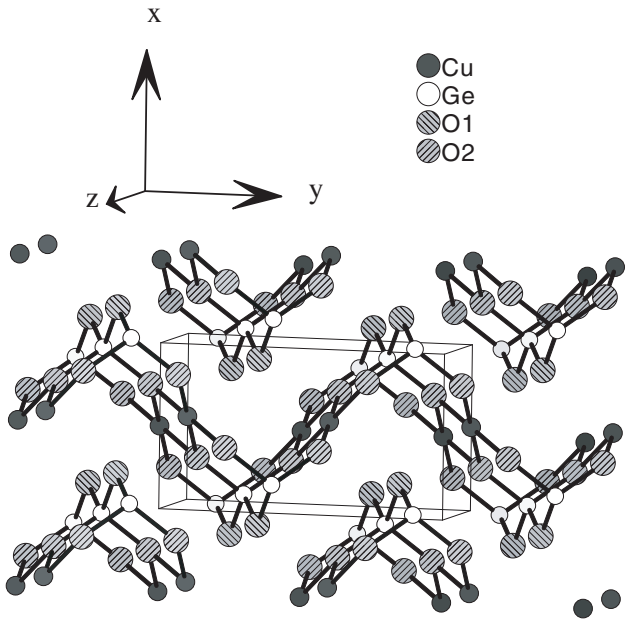}
\caption{ A schematic depiction of the unit cell reference frame we
use for decomposing the polarized spectra.  The unit cell is drawn in
the center of the plot.  Atoms belonging to eight \cugeo\ unit cells
are shown in total.  The symbols O1 and O2 used in the legend refer to
two inequivalent oxygen sites.  Note that in this reference frame one
has $x \| a$, $y \| b$, and $z \| c$, as indicated.
\label{cell}
}
\end{figure}

\begin{figure}
\includegraphics[20mm,2mm][180mm,107mm]{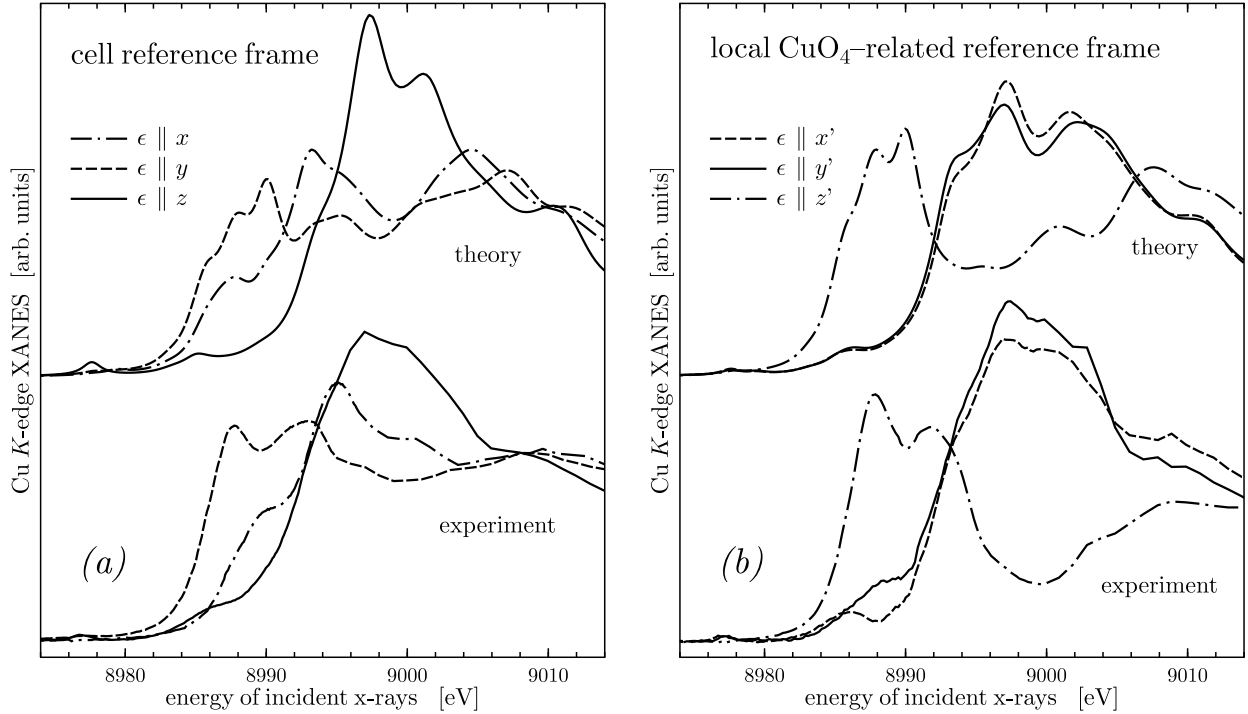}
\caption{ {\it (a)} Experimental (lower graph) and theoretical (upper
graph) Cu \kedge\ polarized XANES partial spectral components $p_{x}$,
$p_{y}$, and $p_{z}$\ resolved in a unit cell reference frame. {\it
(b)} Experimental (lower graph) and theoretical (upper graph) Cu
\kedge\ partial spectral components $p_{x'}$, $p_{y'}$, and $p_{z'}$\
resolved in a local reference frame attached to the nearest Cu
neighborhood.  The theoretical curves were obtained for a cluster of
167 atoms, taking into account a relaxed and screened core hole.
\label{cu}
}
\end{figure}

\begin{figure}
\includegraphics[17mm,2mm][101mm,150mm]{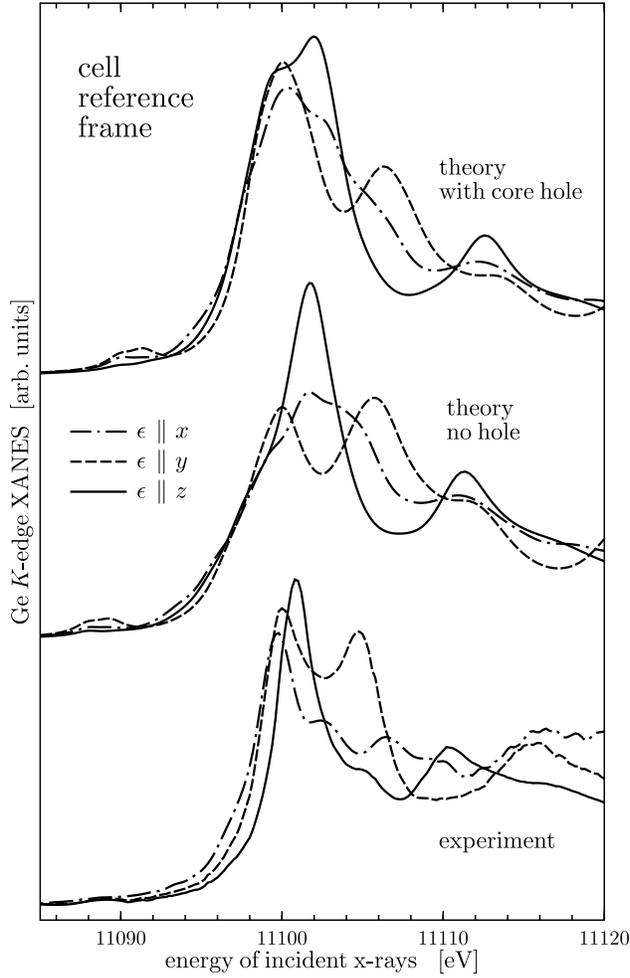}
\caption{ Experimental (lower graph) and theoretical (two upper
graphs) Ge \kedge\ XANES partial spectral components $p_{x}$, $p_{y}$,
and $p_{z}$\ resolved in a unit cell reference frame.  Theoretical
curves obtained for a cluster of 169 atoms are shown both for a ground
state potential (no core hole) as well as for the case when a relaxed
and screened core hole is taken into account.
\label{ge}
}
\end{figure}

By inverting Eq.\ (\ref{polar}) for a triad of measured spectra, the
partial spectral components in a unit cell reference frame can be
extracted (i.e., we take $x \| a$, $y \| b$, and $z \| c$\ --- see
Fig.\ \ref{cell} for a schematic depiction).  The results of such a
decomposition of experimental spectra are shown in the lower graphs of
Fig.\ \ref{cu}{\it a} for the Cu \kedge\ and of Fig.\ \ref{ge} for the
Ge \kedge.  We checked that the partial spectral components obtained
in this way do not depend on the particular choice of the triad of
experimental spectra which was involved in inversion of Eq.\
(\ref{polar}) (see also Refs.\ \onlinecite{v2o5,nasecuo} for a more
thorough discussion).  Our $p_{y}$\ and $p_{z}$\ components of the Cu
\kedge\ XANES are in a good agreement with the corresponding \ebb\ and
\ecc\ components measured by Cruz \ea.\cite{cruz} The remaining
$p_{x}$\ component cannot be measured directly for a crystal cleaved
parallel to the $bc$\ plane and is best accessible through the kind of
decomposition we performed here.

By considering the quadrupole transitions as well, one might be able
to separate the dipole and the quadrupole contributions in the
pre-peaks at 8977~eV and at 11089~eV and even to estimate the angular
character of the quadrupole parts, analogously as it was done for
CuO.\cite{nasecuo} However, the pre-peak intensity is too low in this
case to allow a reliable analysis of this kind.  Therefore, we limit
ourselves to exploring dipole transitions exclusively.

In order to get a better physical insight into the nature of the some
spectral features, one may attempt to perform the spectral
decomposition in other coordinate frames as well.  In particular, this
might be especially useful in a reference frame which reflects the
{\em local} symmetry around the photoabsorbing site.  Although Eq.\
(\ref{polar}) is strictly valid in the unit cell reference frame only,
it has been demonstrated that a consistent resolving of partial
spectral components may be possible in local reference frames as well,
provided that the coordinate axes are chosen in a suitable
way.\cite{nasecuo} One can make an {\it a posteriori} check whether
the coordinate frame was chosen in a suitable way or not by performing
the inversion of Eq.\ (\ref{polar}) for several independent triads of
measured spectra.  Only if the partial spectral components obtained
from different sets of data coincide, the corresponding coordinate
system is a ``good'' one.  The physical interpretation of partial
components in such a local-symmetry-adapted system is often more
transparent than in the case of a unit cell reference frame and may be
more suitable for comparing spectra at absorption edges of atoms of
the same element in different compounds.

\begin{figure}
\includegraphics[70mm,120mm][160mm,180mm]{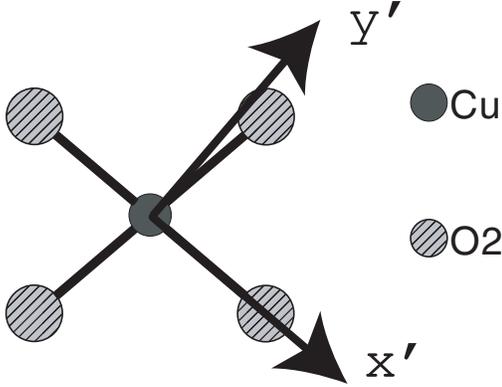}
\caption{ The local reference frame used for resolving partial
spectral components $p_{x'}$, $p_{y'}$, and $p_{z'}$\ in this paper.
The $x'$\ and $y'$\ axes are perpendicular to each other, the $z'$\
axis is perpendicular to the plane of the CuO$_{4}$\ rectangle.
\label{local}
}
\end{figure}

The nearest neighborhood of Cu in \cugeo\ is similar to that of other
copper oxides: Copper is in the center of an oxygen rectangle.  Four
nearest oxygens at a distance of 1.93~\AA\ form its corners, the sides
of the rectangle are 2.50~\AA\ and 2.94~\AA\ long, respectively.  As
can be seen from Fig.\ \ref{cell}, there are actually two
interpenetrating networks of CuO$_{4}$\ rectangles in a \cugeo\
crystal, tilted one to another at an angle of 69$^{\circ}$.  Thus one
has to take into account when inverting Eq.\ (\ref{polar}) that the
measured spectra reflect in fact an average over Cu sites belonging to
two different systems.  A similar situation arises, e.g., in a pure
copper oxide CuO (Ref.\ \onlinecite{nasecuo}).  We have found that the
local partial spectral components $p_{x'}$, $p_{y'}$, and $p_{z'}$\ at
the Cu \kedge\ XANES can be consistently resolved in a local reference
frame where the $z'$\ axis is perpendicular to the CuO$_{4}$\ plane,
the $x'$\ axis coincides with a Cu-O bond and the $y'$\ axis is
perpendicular to both $x'$\ and $z'$\ axes, meaning that it lies in
the CuO$_{4}$\ plane and holds a 9$^{\circ}$\ angle with another Cu-O
bond.  A schematic drawing of the orientation of this local reference
frame with respect to the CuO$_{4}$\ rectangle is shown in Fig.\
\ref{local}.  Partial spectral components resolved in the local
reference frame are displayed in Fig.\ \ref{cu}{\it b}.  It should be
noted that a consistent decomposition of Cu \kedge\ XANES in \cugeo\
cannot be obtained if one chooses the local $x'$\ and $y'$\ axes
parallel to the sides of the CuO$_{4}$\ rectangle (the partial
components derived from different sets of measured spectra do not
coincide in that case).  Hence, the situation is just opposite to the
case of CuO, where the partial components could be consistently
resolved in a local reference frame only if the $x'$, $y'$\ axes were
parallel to the sides of the CuO$_{4}$\ parallelogram.\cite{nasecuo}

We did not attempt to decompose the Ge spectra in a local reference
frame as our primary concern is the Cu edge in relation to spectra of
other copper oxides.

\subsection{Theory}  \label{sec_theory}

Theoretical Cu \kedge\ and Ge \kedge\ partial spectral components were
evaluated by calculating XANES spectra with the polarization vector
\ee\ being pointed to relevant directions.  Self-consistent scattering
potentials were taken over from SCF-$X\alpha$\ molecular
calculations\cite{scf} for clusters of 15--24 atoms, employing an
amended {\sc xascf} code of Case and Cook.\cite{xascf,ag2o} The
results presented in in upper graphs of Figs.\ \ref{cu}--\ref{ge} were
calculated using the real-space multiple-scattering (RS-MS)
formalism,\cite{vvedensky} employing clusters of 167 and 169 atoms for
the Cu and Ge edges, respectively.  These calculations were done using
the {\sc rsms} code, which is an amended descendant of the {\sc
icxanes} code\cite{icxanes} and is maintained by our group.\cite{rsms}
A few more technical details about our implementation of the RS-MS
technique can be found elsewhere.\cite{nasecuo,druhecuo} In order to
verify the robustness of our results, we compared the results obtained
by our codes and by the selfconsistent {\sc feff8} code\cite{feff}
(version 8.10). Both code packages rely on a bit different
implementations of the one-electron multiple-scattering formalism:
E.g., the interstitial region is suppressed in the stage of finding
the selfconsistent potential in the {\sc feff8} code while we use a
proper muffin-tin formalism and Norman muffin-tin sphere radii are
employed in {\sc feff8} while we use matching-potential radii. Despite
these differences, both codes gave similar results for identical
clusters (we used 95 atoms for Cu edge and 91 atoms for Ge edge
spectra in this comparative study).  The main difference consisted in
up to 20\% lower intensities of some low-energy (\mm{E \lesssim
25}~eV) spectral peaks for the {\sc feff8} code, giving thus worse
agreement with experiment than our {\sc xascf} and {\sc rsms} codes.
We did not try to optimize {\sc feff8} input parameters to identify
the source of this difference, our goal has been just to check that
the two codes do not lead to contradictory pictures.  A comparison of
XANES spectra calculated for various settings of the muffin-tin
potential parameters can be found, e.g., in Ref.\ \onlinecite{v2o5}.

The XANES calculations for 167/169-atoms clusters were performed both
without any core hole and with a relaxed and screened core hole taken
into account (i.e., a 1$s$\ electron was removed from the central atom
and put among the valence electrons and only thereafter the electronic
structure of the cluster was selfconsistently calculated).
Neither choice leads to essentially better agreement between the
theory and experiment than the other one.  Generally, the core hole
increases the intensity of peaks at the low-energy part of the
spectra.  This effect is not significant at the Cu edge so we display
only the results of the calculation which takes the core hole into
consideration in Fig.\ \ref{cu} (the intensities of theoretical peaks
at 8993~eV for the $p_{y}$\ component and at 8997~eV for the $p_{z}$\
component would be about 10\% lower if the core hole was neglected).
At the Ge edge the core hole effect is more significant so we display
both sets of curves in Fig.\ \ref{ge}.  It appears that including the
core hole actually ``overcorrects'' the ground state results in this
case and consequently does not give rise to distinctly better agreement
between theory and experiment for the Ge edge XANES.

We made analogous calculations for non-selfconsistent potentials
constructed according to the Mattheiss prescription (superposition of
potentials and charge densities of isolated atoms), too.  As could be
anticipated, selfconsistent potentials give rise to theoretical
spectra which agree better with experiment than in the case of
non-selfconsistent potentials, however, the improvement is not an
essential one.  All distinct spectral features that can be observed at
the theoretical curves in Figs.\ \ref{cu}--\ref{ge} are present also
in corresponding spectra obtained for non-selfconsistent potentials,
with peak positions and intensities differing no more than
$\approx$20\% (we do not show those curves for brevity).  Thus use of
self-consistent potentials is not crucial for our study.

The alignment in energy of the theoretical curves shown in Figs.\
\ref{cu}--\ref{ge} was set so that the best overall agreement between
theory and experiment is achieved.  All theoretical curves were
convoluted with a Lorentzian function of the width $w$\ given by the
{\it ansatz} \mm{w = w_{c} + 0.03\!\times\! E}, where the constant
part $w_{c}$\ accounts for the core hole lifetime\cite{lifetime} and
the energy-dependent part mimics the finite lifetime of a
photoelectron with an energy $E$.  The factor 0.03 was chosen just by
convenience --- we did not attempt to achieve the best possible
agreement between theory and experiment by optimizing the smearing
function.

\section{Discussion}

Let us concentrate first on the Cu \kedge\ in the unit cell reference
frame, where a partial comparison with the work of Cruz \ea\cite{cruz}
is possible.  As it can be seen from Fig.\ \ref{cu}{\it a}, our
calculation describes the gross shape of the three components $p_{x}$,
$p_{y}$, $p_{z}$\ and their polarization splitting.  However, only the
$p_{z}$\ component, probing states parallel to the $c$\ axis, is
reproduced by the theory as concerns the number of features (shoulders
at 8986~eV and 8993~eV and peaks at 8997~eV, 9000~eV, and 9008~eV at
the experimental curve), their positions (within 1--2~eV's), and
relative intensities.  The reason why the two peaks of the main
doublet in $p_{z}$\ are not clearly separated in the experiment
(either ours or that of Cruz \ea\cite{cruz}) is not clear.  It may be
a consequence of experimental energy resolution (most probably arising
from the so called thickness effect --- absorption maxima are
suppressed and consequently smeared for thick samples) or it may stem
from a smearing caused by many-body effects, not accounted for by our
theory.  The other two experimental components $p_{x}$, $p_{y}$\ seem
to include an ingredient which defies our theoretical description
substantially more than is the case of the $p_{z}$\ component.

Cruz \ea\cite{cruz} interpret the Cu \kedge\ XANES of \cugeo\ in this
energy range by semiquantitative arguments, relying on comparison with
Cu spectra of \lacuo\ and CuO.  They suggest an essentially many-body
interpretation of dominant spectral features: The double structure of
the main peaks at 8987~eV and 8993~eV for the $p_{y}$\ component and
at 8997~eV and 9000~eV for the $p_{z}$\ component (see Fig.\
\ref{cu}{\it a}) is ascribed to transitions to well-screened final
states followed by satellite transitions to poorly screened states.
The success of our one-electron LDA-type computation to describe at
least the $p_{z}$\ component correctly seems to contradict this point
of view.  Of course, no definite conclusion regarding the one-electron
or many-body nature of the main $p_{z}$\ peak splitting can be made
until a proper many-body calculation is done.

As noted above, the experimental $p_{x}$\ and $p_{y}$\ components in
Fig.\ \ref{cu}{\it a} display more serious deviations from the
one-electron theory than the $p_{z}$\ component.  In order to identify
the source of these deviations, it might be useful to explore the
partial spectral components in the local reference frame.  Indeed, if
we compare the theoretical and experimental curves in Fig.\
\ref{cu}{\it b}, it emerges clearly that the primary source of
discrepancy between the theory and the experiment lies mainly {\em
outside the} CuO$_{4}$\ {\em plane}.  The peak positions and relative
intensities of the $p_{x'}$\ and $p_{y'}$\ components are reproduced
by theory with a similar accuracy as was the case with the $p_{z}$\
component in the cell reference frame.  The absence in the experiment
of a well-defined doublet structure of the main peak at 8997--9004~eV
is probably caused by the same mechanism as a similar situation with
the cell $p_{z}$\ component mentioned above.  The local $p_{z'}$\
component is described by the theory significantly worse than the
local $p_{x'}$\ and $p_{y'}$\ components, especially as concerns the
fine structure of the peak at 8988--8992~eV in Fig.\ \ref{cu}{\it b}.

Having this in mind, one can re-assess the situation in the unit cell
reference frame: As the $c$\ axis runs parallel to the CuO$_{4}$\
plane, the $p_{z}$\ component probes solely states lying within this
plane and, hence, can be correctly characterized by the theory (cf.\
Fig.\ \ref{cell}).  On the contrary, the $p_{x}$\ and $p_{y}$\
components probe states lying within the CuO$_{4}$\ plane as well as
states reaching perpendicular to it.  It is the
perpendicular-to-CuO$_{4}$\ admixture into these components which
spoils the agreement between theory and experiment.  Without this
mixing, one could have expected for Fig.\ \ref{cu}{\it a} a similar
agreement between the theory and experiment as can be seen in Fig.\
\ref{cu}{\it b} for the $p_{x'}$\ and $p_{y'}$\ components.

The failure of the theory to describe the $p_{z'}$\ component in the
local reference frame is not a total one: The theory accounts for the
general trends of the $p_{z'}$\ spectral curve such as existence of a
well-separated maximum at 8988--8992~eV followed by a broad structure
around 9008~eV.  The most notable deficiency of the theory for this
polarization is not reproducing the double-peak structure at 8988~eV
and 8992~eV. This failure then transforms into deficiencies in
describing the unit cell $p_{x}$\ and $p_{y}$\ components in this
energy region.  Secondly, the fine structure of the second broad
feature (positions and relative intensities of sub-maxima at 9003~eV
and 9008~eV) is not correctly reproduced by our calculation.
 
At this point, it is difficult to identify the reason
of this deficiency of the theory more specifically.  One could make
use of the analogy with \lacuo, where the theory also describes the
states within the Cu-O layers fairly accurately but fails in
describing XANES components perpendicular to that layer.\cite{guo,li}
Guo \ea\cite{guo} attribute this failure to the presence of shake-up
satellite peaks in \lacuo\ spectra and present plausible arguments for
their view.  However, the situation may be different for \cugeo: Among
others, the failure of the theory for the out-of-plane XANES
components appears to be much more dramatic for \lacuo\ than for
\cugeo\ (a whole peak missing in \lacuo\ spectrum vers.\ an inverted
double-peak intensity in \cugeo\ spectrum).  Since the calculations of
\lacuo\ spectra\cite{guo,li} as well as our RS-MS calculation of
\cugeo\ XANES rely on the muffin-tin approximation, one cannot rule
out full-potential effects as the main culprits.  Intuitively one
would expect non-muffin-tin corrections to be more important for the
more loose out-of-plane states than for the states in the more
closely-packed plane of CuO$_{4}$\ rectangles, so this explanation
would be in accord with our results.  Anyway this question seems to
remain open.

As can be seen from Fig.\ \ref{cu}, the pre-peak at the Cu \kedge\
appears to be reproduced at the correct position even if quadrupole
transitions are neglected in the calculation.  We verified that if the
quadrupole transitions are included, they are completely negligible in
the whole energy range except for the pre-edge region around 8977~eV,
where they contribute with approximately with the same intensity as
the dipole transitions.  This might serve as an indication of a mixed
dipole-quadrupole nature of the pre-peak at the Cu \kedge\ of \cugeo,
however, one should be quite cautious about this issue as the
successful reproduction of the pre-peak position by our calculation
may be just a fortuitous coincidence: The pre-edge XANES intensity
crucially depends on the energy position of the onset of unoccupied
states and on the structure of electronic states very close to this
onset.  This kind of information cannot be reliably obtained via an
LDA-type calculation for a strongly-correlated material such as
\cugeo\ (see Refs.\ \onlinecite{mattheiss_cugeo,popovic} for LDA
calculations of electronic structure of \cugeo\ and Refs.\
\onlinecite{atzkern,zagoulaev} for calculations going beyond LDA).
Note that, e.g., in the case of CuO the agreement between theory and
experiment in the pre-edge region is seemingly better for a
non-selfconsistent potential than for a self-consistent
one.\cite{druhecuo} So we think that we cannot draw really reliable
conclusions regarding the pre-edge region and, consequently, do not
even display the corresponding ``quadrupole--included'' curves here.

The Ge \kedge\ is decomposed in the unit cell reference frame in Fig.\
\ref{ge}.  As noted already in Sec.\ \ref{sec_theory}, the best
agreement between theory and experiment would be attained for curves
lying somewhat between the data obtained when the core hole is present
and when it is neglected.  Both calculations account for the strong
polarization (i.e., angular) dependence of the experimental curves
fairly well.  However, not all components are described by the theory
with the same accuracy.  The main peak of the $p_{x}$\ component is
too broad in the calculated spectra and neither the fine oscillations
seen in the experimental $p_{x}$\ curve are reproduced by the theory
appropriately (though including the core hole improves this a bit).
The main doublet in the $p_{y}$\ component is reproduced, however, the
separation of its peaks is 5~eV in the experiment and 6~eV in the
theory.  At the $p_{z}$\ component, the calculation without the core
hole reproduces the 11101~eV and 11110~eV experimental peaks both in
positions and relative intensities, just the 11105~eV shoulder is
missing in the calculated curve.  If the core hole is included, the
low-energy shoulder at 11099~eV of the theoretical $p_{z}$\ curve is
drastically increased, worsening thus the agreement with experiment.
It seems therefore that a proper description of the core hole effect
at the Ge edge of \cugeo\ has to go beyond the static relaxed and
screened model. We did not try to employ other simple --- less
frequently used --- prescriptions for core hole treatment (unrelaxed
and/or unscreened hole) as they have not worked well with other
non-metallic systems.\cite{cugase}

Overall it appears that the best agreement between theory and
experiment has been achieved for the \ecc\ component $p_{z}$, next
comes the \ebb\ curve and, finally, worst agreement occurs for the
\eaa\ component $p_{x}$.  Although we are unable to find unambiguously
what is the reason for this ``hierarchical'' behavior, we believe that
the degree of packing of the neighborhood of Ge atoms in various
directions may be a factor.  Namely, the Ge atoms are nearly
tetrahedrally coordinated by four O atoms at 1.74--1.79~\AA\ and the
$a$, $b$, and $c$\ directions hold different angles with those
oxygens.  The $a$\ direction avoids the tetrahedron corners as much as
possible, holding a 54$^{\circ}$\ angle with all the Ge-O bonds.  The
minimum angle between the $b$\ and $c$\ directions and any of the Ge-O
bonds is 36$^{\circ}$, on the other hand.  Moreover, the $c$\
direction points exactly to the Ge atoms which form the second
coordination shell around germanium at 2.94~\AA.  Hence, a rule of
thumb emerges that the more open is the local geometry probed by a
particular spectral component $p_{x}$, $p_{y}$, or $p_{z}$, the worse
agreement between theory and experiment at the Ge \kedge\ arises.  A
possible explanation of this correlation might rest with the
muffin-tin approximation we employ, as non muffin-tin effects are
expected to be more significant at open geometries than at close ones.

Experimental XANES spectra at the O \kedge\ of \cugeo\ were presented
by Corradini \ea\cite{corradini} and by Agui \ea\cite{agui} for
energies up to 25~eV above the absorption threshold. The
interpretation of O edge spectra is complicated by the fact that there
are two inequivalent oxygen sites in \cugeo, meaning that one sees a
superposition of two edges in the experiment actually.  We attempted
to reproduce the polarization dependence of the O \kedge\ spectra
measured by Corradini \ea\cite{corradini} by a RS-MS calculation for
the same self-consistent potential which we employed for calculating
the curves displayed in Figs.\ \ref{cu}--\ref{ge}.  The energy shift
between the edges of the two inequivalent oxygens was set by hand, as
there are several factors affecting its value which cannot be
described with a sufficient accuracy by an LDA-type calculation (a
brief discussion of those effects can be found, e.g., in Ref.\
\onlinecite{corradini}).  We have found that there is no suitable
value of such a shift which would lead to the reproduction of the two
distinct polarization-dependent peaks within the first 3~eV above the
O \kedge\ threshold, as observed in the experiment.\cite{corradini} On
the other hand, the relatively broad main peak at 5--10~eV above the
threshold does not exhibit any distinct angular-dependent features at
all and, consequently, can be at least roughly reproduced for several
different values of the energy shift between the two O edges, offering
thus no clue for its proper value.  We do not display the results here
for brevity, concluding that our calculation cannot be used as a guide
for interpreting the polarization dependence of the O \kedge\
absorption spectrum of \cugeo.  Strong electron correlations,
unaccounted for by the LDA framework, seem to be the most probable
explanation for the failure of our calculations to reproduce the
polarized O \kedge\ XANES close to the threshold.  Note that for that
reason (unreliability of LDA at the x-ray absorption threshold in
\cugeo) we do not discuss the pre-peaks at the Cu and Ge edges in
detail either.  On the other hand, those parts of the spectra which
correspond to photoelectron energies higher than \mm{E \gtrsim 5}~eV
can be at least partially described by a one-electron LDA-type
approach, as demonstrated above (Figs.\ \ref{cu}--\ref{ge}).

The fact that a relatively simple theory appears to be able to
reproduce gross trends in XANES of an insulating, probably
charge-transfer oxide is quite surprising on its own. It seems that
strong electron correlations in this type of compound need not be so
crucial at photoelectron energies {\em above the pre-edge}, contrary
to what has been generally anticipated so far.

\section{Conclusions}

The main features of polarized Cu \kedge\ and Ge \kedge\ XANES of
\cugeo\ can be described fairly well within the one-electron
framework.  Contrary to earlier suggestions,\cite{cruz} the main peak
splitting for the \ecc\ polarization at the Cu \kedge\ can be
correctly reproduced within the one-electron LDA-type approach (no
need for superpositioning of well-screened and poorly-screened
states).  The source of the remaining discrepancies between theory and
experiment at the Cu \kedge\ XANES can be better identified and
understood if the polarized spectra are decomposed in the local
reference frame attached to the nearest Cu neighborhood rather than in
the unit cell reference frame.  We have found that it is the
out-of-CuO$_{4}$-plane $p_{z'}$\ component which defies the
one-electron RS-MS calculation for a muffin-tin potential.  For the Ge
\kedge\ XANES, the agreement between the theory and the experiment
appears to be better for those polarization components which probe
more compact local surroundings than for those which are directed to
the interstitial parts of the crystal. Including the core hole affects
teh Ge edge theoretical spectra significantly more than at the Cu edge
and, to a certain degree, leads to a worse agreement with Ge edge
experiment in comparison with the case when the core hole is
neglected.

\section*{Acknowledgements}

The theoretical part of this work was supported by grant 202/02/0841
of the Grant Agency of the Czech Republic.  The experimental part was
supported by Hamburger Synchrotronstrahlungslabor HASYLAB Project No.\
II-98-058.  The use of the {\sc crystin} structural
database\cite{crystin} was financed by grant 203/02/0436 of the Grant
Agency of the Czech Republic.

\end{document}